\documentclass[10pt]{article}
\usepackage[OE]{express}

\begin{document}

\title{Approaching perfect absorption of monolayer molybdenum disulfide at visible wavelengths using critical coupling}

\author{Xiaoyun Jiang, Tao Wang,\authormark{*} Shuyuan Xiao, Xicheng Yan, Le Cheng, and Qingfang Zhong}

\address{Wuhan National Laboratory for Optoelectronics, Huazhong University of Science and Technology, Wuhan 430074, China}
\email{\authormark{*}wangtao@hust.edu.cn} 



\begin{abstract}
A simple perfect absorption structure is proposed to achieve the high efficiency light absorption of monolayer molybdenum disulfide (MoS$_{2}$) by the critical coupling mechanism of guided resonances. The results of numerical simulation and theoretical analysis show that the light absorption in this atomically thin layer can be as high as $98.3\%$ at the visible wavelengths, which is over $12$ times more than that of a bare monolayer MoS$_{2}$. In addition, the operating wavelength can be tuned flexibly by adjusting the radius of the air hole and the thickness of the dielectric layers, which is of great practical significance to improve the efficiency and selectivity of the absorption in monolayer MoS$_{2}$. The novel idea of using critical coupling to enhance the light-MoS$_{2}$ interaction can be also adopted in other atomically thin materials. And the meaningful improvement and tunability of the absorption in monolayer MoS$_{2}$ provide a good prospect for the realization of high-performance MoS$_{2}$-based optoelectronic applications, such as photodetection and photoluminescence.
\end{abstract}



\section{Introduction}
Atomically thin two-dimensional ($2$D) materials, such as graphene and monolayer transition metal dichalcogenides (TMDCs), have been intensively studied due to their remarkable mechanical, electrical and optical properties\cite{bonaccorso2010graphene,wang2012electronics,he2016further}. The unique characteristics of graphene contain the ultra-wide spectral response, the ultra-high carrier mobility and dynamic tunability, which open up many exciting possibilities for nano-integrated optoelectronic devices such as optical detectors\cite{zhang2015towards,xiao2016tunable}, modulators\cite{yan2017dynamically,xiao2018active}, metamaterial perfect absorbers (MPA)\cite{jiang2017tunable,xia2017multi}, and biological or chemical sensors\cite{amin2013dynamically,yan2017high}. However, pure graphene, though extremely promising, has a fixed zero band gap, which makes it uninteresting from the point of view of the device application. In contrast, TMDCs, a new class of semiconductors with distinctive properties at a single layer thickness, has attracted widespread attention because of its interesting and important optoelectronic characteristics\cite{wang2012electronics}. In particular, molybdenum disulfide (MoS$_{2}$), representing one of the most special candidates for TMDCs semiconductor applications, has a direct bandgap (around $1.8$ev) for electronic transition when the bulk structure is converted into monolayer\cite{bahauddin2016broadband}. Owing to its special direct band gap and low dark current, monolayer MoS$_{2}$ has been considered as the ideal atomically thin $2$D material for photoluminescence\cite{sobhani2014enhancing}, photodetection\cite{lopez2013ultrasensitive}, and photovoltaic devices\cite{pospischil2014solar,li2017total}. It is well known that the maximum absorption of a thin absorbing layer under a single illumination in the air cannot exceed $50\%$\cite{botten1997periodic}, and the inherent thickness of the atomically thin layer also hinders the better interaction between light and matter. Therefore, the light absorption in monolayer MoS$_{2}$ is weak, which seriously restricts the application of MoS$_{2}$ in the optoelectronics area. For instance, the sing-pass absorption of monolayer MoS$_{2}$ is only $10\%$ for an atomically thin layer (${\sim} 0.615$ nm thickness) in the visible range\cite{janisch2016mos2}. So in order to achieve many high-performance MoS$_{2}$-based optical devices, it is highly desirable to enhance absorption up to $100\%$. Recently, several approaches have been proposed to improve light absorption in monolayer MoS$_{2}$\cite{lu2017nearly,piper2016broadband,long2017magnetic,butun2015enhanced}. A new multilayer photonic structure was proposed to realize the absolute absorption of light in monolayer MoS$_{2}$ reached up to $96\%$ in the visible range due to the excitation of highly-confined Tamm plasmon mode\cite{lu2017nearly}. Through the consideration of light trapping theory, the structure of photonic crystal slab as a resonator was designed, in which ideal metallic reflector or dielectric Bragg reflector was considered as a perfect mirror. The optimized design parameters were used to achieve average absorption of $51\%$ over the entire visible spectrum in monolayer MoS$_{2}$ under normal incidence\cite{piper2016broadband}. The magnetic coupling metasurface has been investigated to boost the light absorption of single MoS$_{2}$ layer, which could reach $72.7\%$ within the visible wavelength range \cite{long2017magnetic}. In addition, the strong plasmonic resonances in metallic nanostructures have been widely utilized to boost the light absorption of MoS$_{2}$\cite{butun2015enhanced}. However, up to now, for monolayer MoS$_{2}$ with atomically thin layer, a much simpler structure with the dynamical wavelength-selective property and ultra-high-efficiency light absorption has yet to be reported.
 
In this article, we investigate ultra-high-efficiency light absorption in monolayer MoS$_{2}$ by critical coupling with guided resonance, in the band gap region of MoS$_{2}$, in which the MoS$_{2}$ layer is coupled to a $2$D periodic dielectric configuration with low refractive index contrast. Compared with previous research results, we propose a much simpler MPA to effectively enhance the light-matter interaction, resulting in the light absorption of monolayer MoS$_{2}$ up to $98.3\%$. Meanwhile, we are able to control absorption efficiency of MoS$_{2}$ at different operating wavelengths by adjusting the thickness of dielectric layers and the radius of air hole. In addition, the same principle can be extended to other atomically thin $2$D materials, such as MoSe$_{2}$, WSe$_{2}$ and WS$_{2}$. And we also propose a promising way to enhance the interaction of light with these $2$D materials, which are very important for the application of TMDC-based photodetection and photoluminescence.

\section{Structure and model}
\begin{figure}[htbp]
\centering
\includegraphics[scale=0.4]{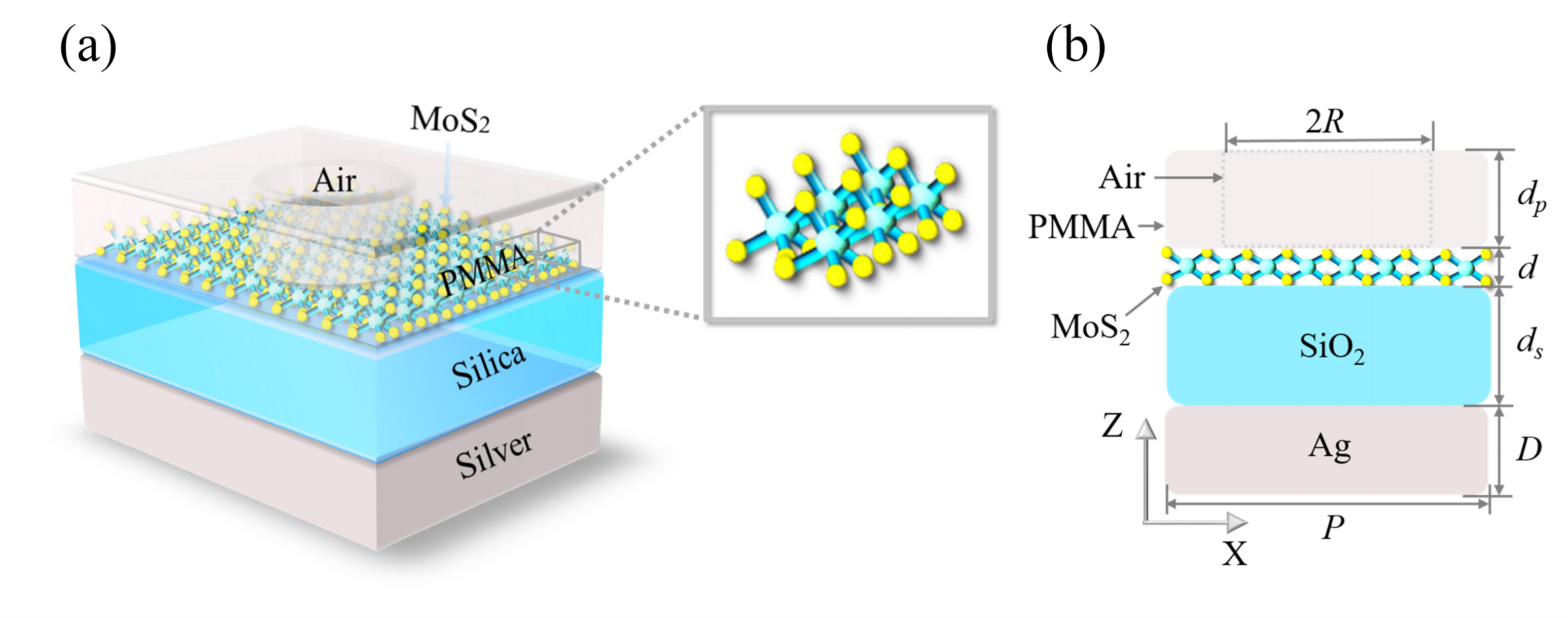}
\caption{(a) $3$D schematic illustration of absorption enhancement system for monolayer MoS$_{2}$. The system consists of a metallic mirror (gray) and a multilayer dielectric combinations, in which the monolayer MoS$_{2}$ is the shape of honeycomb fishnet inserted in the middle of multilayer dielectric. The enlarged illustration represent the structure of the atomically thin MoS$_{2}$ layer. (b) Cross-sectional view of the designed structure. Here, \textit{$d_{p}$}, \textit{$d$}, \textit{$d_{s}$}, and \textit{$D$} stand for the thickness of PMMA, MoS$_{2}$, SiO$_{2}$, and Ag, respectively. \textit{$R$} and \textit{$P$} represent air hole radius and the lattice period, respectively. The position of the dotted line stands for the air hole.}
\end{figure}
Figure. 1(a) shows the schematic of the proposed monolayer MoS$_{2}$-based perfect absorber. A monolayer MoS$_{2}$ is sandwiched between a silica (SiO$_{2}$) layer and a $2$D polynethy $1$-methacrylate (PMMA) with periodic air holes, and a silver (Ag) layer is deposited on the back side of SiO$_{2}$ layer as a mirror to block the transmission of light \cite{barrios2004compact}. Numerical simulations are analyzed by using finite-difference time-domain (FDTD) method-based software from Lumerical Solutions. In the simulations, the perfectly matched layers (PML) absorbing boundary conditions are adopted at the bottom and top of computational domain, while the periodic boundary conditions are set on the left and right sides of the model. First of all, we consider that the TM-polarized (the electrical polarization parallel to the X-axis) light project on the structure under normal incidence. The x-z cross section of the designed structure is described in Fig. 1(b). The refractive indices of air, PMMA, and SiO$_{2}$ are taken to be $1$, $1.48$, and $1.45$, respectively. The material of the metallic mirror is chosen as Ag, whose relative permittivity in the wavelengths of interest can be given by the Drude model: $\varepsilon(\omega) = \varepsilon_{\infty} - {\omega_p^2} / {(\omega^2 + i \gamma \omega)}$. Here, for Ag, these physical parameter can be configured as $\varepsilon_{\infty} = 3.7$, $\omega_p = 1.38\times 10^{16}$ s$^{-1}$, and $\gamma = 2.73\times 10^{13}$ s$^{-1}$ \cite{lu2017nearly,lu2016tunable,cheng2017polarization}. In numerical calculations, the wavelength-dependent complex dielectric constant of monolayer MoS$_{2}$ with a thin thickness of $d=0.615$ nm measured experimentally by Li \textit{et al}. is employed in our simulations\cite{li2014measurement}. Meanwhile, the thickness of PMMA, SiO$_{2}$, and Ag layers are set as $d_{p}=205$ nm, $d_{s}=255$ nm, and $D=200$ nm, respectively. And the lattice period (\textit{$P$}) and the radius of the air hole (\textit{$R$}) are taken to be $480$ nm and $190$ nm, respectively. The relevant geometric parameters are shown in Fig. 1.

\section{Results and analysis}
\begin{figure}[htbp]
\centering
\includegraphics[scale=0.31]{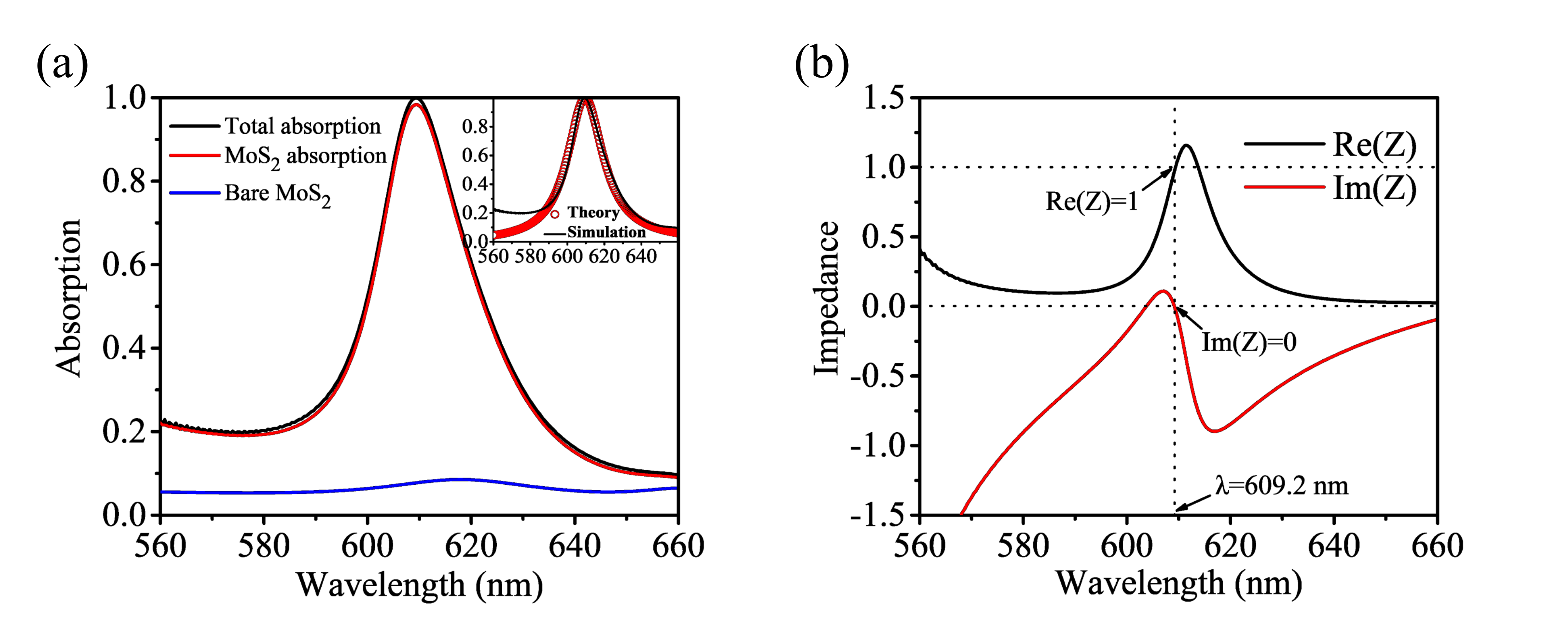}
\caption{(a) Absorption spectrum of a MoS$_{2}$ monolayer (red line) in a MPA at $P = 480$ nm, $R = 190$ nm, $d_p = 205$ nm, $d_s = 255$ nm, $d = 0.615$ nm, and $D = 200$ nm. The light absorption spectra of the whole MPA (black line) and a bare MoS$_{2}$ monolayer (blue line) are shown for comparison. Inset shows that FDTD-simulated (black line) and CMT-fitted (red circle) absorption spectrum in the critical coupling state. Normal incident light is TM-polarized. (b) Real (black line) and imaginary (red line) parts of the effective impedance ($Z$) of the perfect absorption peak are also calculated. The dashed lines represent the free space impedance ($Z_0=1$) and the position of the resonance wavelength, respectively.}
\end{figure}
It is well know that in order to enhance absorption, the common method is to use resonances, but in the visible range, the monolayer TMDCs do not have strong resonant behavior. Therefore, the TMDC layer is inserted into the designed resonant structure to take advantage of the resonance in the structure. In addition, the atomically thin monolayer MoS$_{2}$ has a high light transmission in the visible range, which causes minimal interference to the behavior of the resonator. As shown in Fig. 2(a), simulation results show that the calculated absorption spectra of the entire MPA and the contribution from MoS$_{2}$ at the normal TM-polarized light, where $P = 480$ nm, $R = 190$ nm, $d_p = 205$ nm, and $d_s = 255$ nm, respectively. Simultaneously, the corresponding spectra of a bare MoS$_{2}$ suspended in air also is given for a better comparison. It is found that the incident light is completely absorbed by the whole MPA (black line) at the resonant wavelength of 609.2 nm, among which about $98.3\%$ absorption (red line) occurs in MoS$_{2}$. And the light absorption is improved by approximately $12$ times more than that of a bare MoS$_{2}$ monolayer (blue line). In order to theoretically verify the numerical results, the coupled mode theory (CMT) can be used to describe both enhancement and suppression of light absorption\cite{piper2014total}. In addition, for our chosen subwavelength structure, only the zero-order guided mode resonance can be excited and all other higher orders will not exist, which corresponds to a sharp peak in the absorption spectrum in Fig. 2(a). We assume a resonator, as shown in Fig. 1, has the amplitude of the input and output waves of $u$ and $y$ at a single resonance at $\omega_{0}$, respectively. The external leakage rate of the resonator and the inherent loss rate of monolayer MoS$_{2}$ are assigned to $\gamma_{e}$ and $\delta$, respectively, and the reflectivity of the entire system can be approximately described by the following equation\cite{jiang2017tunable},
\begin{equation}
\Gamma (\omega) \equiv \frac{y}{u} = \frac{j(\omega - \omega_0) + \delta - \gamma_e}{j(\omega - \omega_0) + \delta + \gamma_e},
\end{equation}
and light absorption in a MPA is then calculated as
\begin{equation}
A(\omega) = 1-|\Gamma (\omega)|^2
= \frac{4 \delta \gamma_e}{(\omega - \omega_0)^2 + (\delta + \gamma_e)^2}.
\end{equation}
From the above equations, we note that when the system is in the state of resonance ($\omega = \omega_0$), and the external leakage rate is the same as the inherent loss rate ($\gamma_e = \delta$), which corresponds to the critical coupling state of the system. At this point, the reflection coefficient of a whole MPA disappears and all incident light is absorbed. As shown in the inset of Fig. 2(a), in the vicinity of the resonance, the numerical simulations (FDTD-simulated) are in agreement with the theoretical calculations (CMT-fitted) under the critical coupling condition. However, we note that there are some deviations between the simulation and the theory at the far from the resonant wavelength region because the theory assumes a lossless resonator at this time. On the other hand, from the point of view of macroscopic electromagnetism, when the critical coupling condition is satisfied, the impedance of the whole structure should be equal to that of the free space ($Z_0 = 1$). The effective impedance ($Z$) of the entire MPA can be expressed as\cite{smith2005electromagnetic}
\begin{equation}
Z = \frac{(T_{22} - T_{11}) \pm \sqrt[]{(T_{22}-T_{11})^2 + 4T_{12}T_{21}}}{2T_{21}}.
\end{equation}
The values of the $T$-matrix elements can be calculated from the $S$-matrix elements as well:
\begin{equation}
T_{11} = \frac{(1 + S_{11})(1 - S_{22}) + S_{21}S_{12}}{2S_{21}},
\end{equation}
\begin{equation}
T_{12} = \frac{(1 + S_{11})(1 + S_{22}) - S_{21}S_{12}}{2S_{21}},
\end{equation}
\begin{equation}
T_{21} = \frac{(1 - S_{11})(1 - S_{22}) - S_{21}S_{12}}{2S_{21}},
\end{equation}
\begin{equation}
T_{22} = \frac{(1 - S_{11})(1 + S_{22}) + S_{21}S_{12}}{2S_{21}}.
\end{equation}
Where $S$ and $T$ represent the scattering ($S$) matrix and the transfer ($T$) matrix of the whole structure, respectively. And the two roots of the effective impedance ($Z$) equation correspond to the two paths of light propagation, where we take the plus sign to represent the positive direction in Eq. (3). As shown in Fig. 2(b), at the resonance wavelength of 609.2 nm, the effective impedance of the structure perfectly matches the free space impedance ($Z_{\omega_0} = 1$), achieving the perfect absorption of the entire structure, in which the basic principle is to minimize the reflectivity through impedance matching with free space ($Z_0 = 1$), and to eliminate the light transmittance by a metallic mirror with a certain thickness. In other words, when the incident light is coupled with the guided mode resonance of the designed structure, in the vicinity of the resonant wavelength, the electric field intensity near the monolayer MoS$_{2}$ is obviously enhanced, which significantly boosting the light-matter interaction, and the light absorption in MoS$_{2}$ can be greatly improved. In addition, under the critical coupling condition, the light absorption in monolayer MoS$_{2}$ can reach the maximum value, and all incident light is completely absorbed by the whole structure at the same time. As depicted in Fig. 3(a) and 3(b), when the resonant cavity is in on-resonant ($\lambda = 609.2$ nm) state, the electric field ($|E|^2$) near the monolayer MoS$_{2}$ is strongly enhanced under the critical coupling condition as shown in Fig. 3(a). On the contrary, when the resonant cavity is in off-resonant (namely, $\lambda = 660$ nm) state, the reflection coefficient of a lossless system can be equivalent to -1, and the distribution of electric field intensity is shown in Fig. 3(b).
\begin{figure}[htbp]
\centering
\includegraphics[scale=0.32]{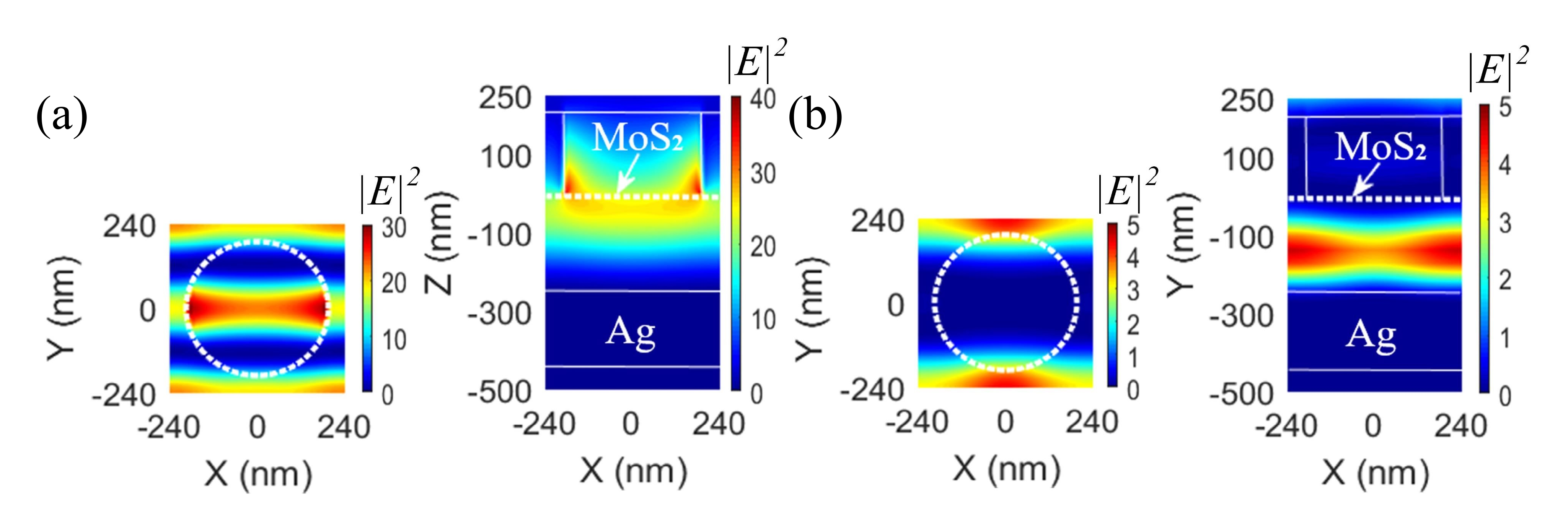}
\caption{(a)Electric field amplitude $|E|^2$ distributions of the proposed structure with on-resonant wavelength (a) and off-resonant wavelength (b). MoS$_{2}$ is located at $z = 0$, corresponding to the dashed lines position in the graph, and the dashed circles represent the position of the air hole. The white lines outline the profile of different materials.}
\end{figure}

\begin{figure}[htbp]
\centering
\includegraphics[scale=0.26]{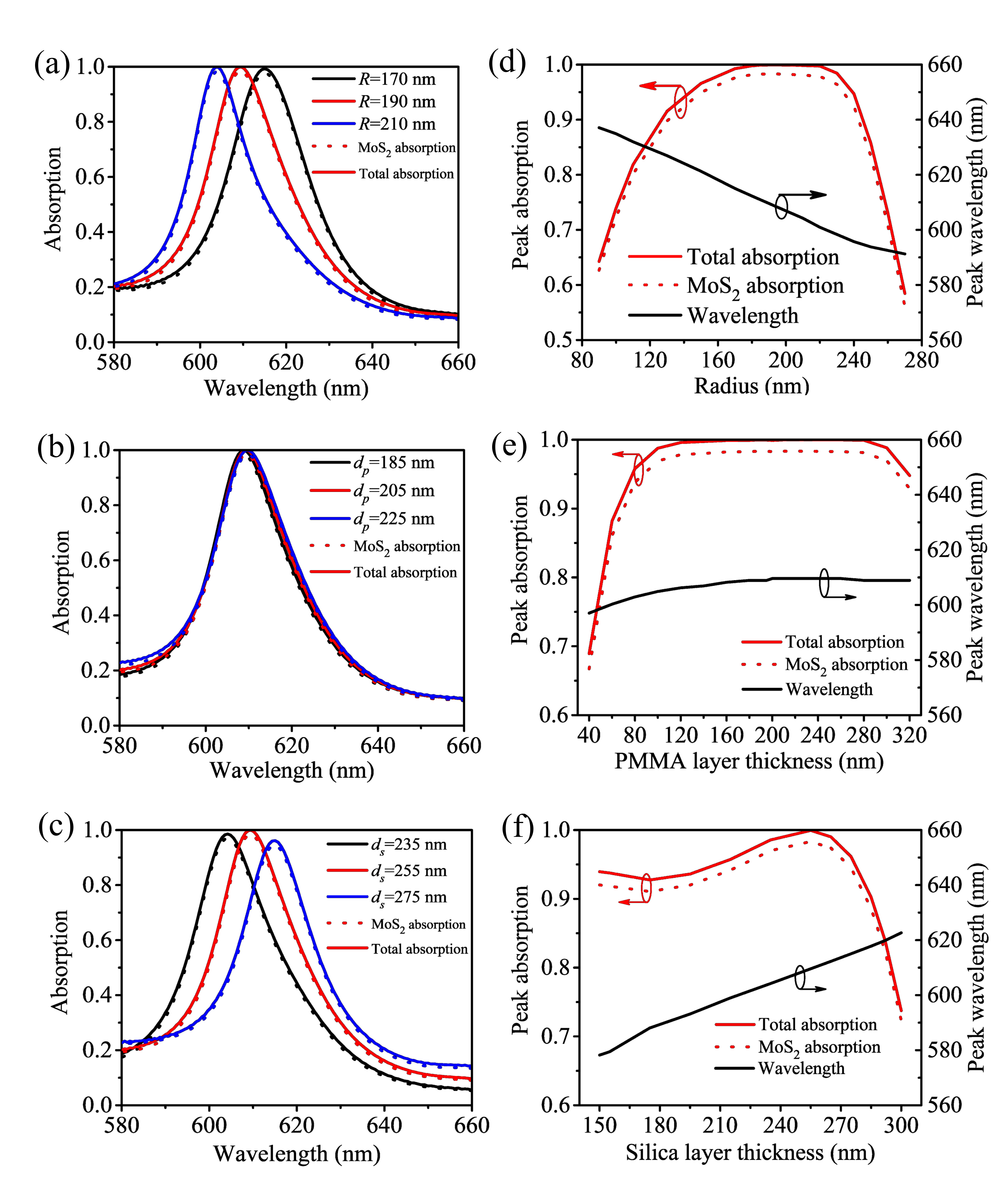}
\caption{The absorption spectra in MPAs. The whole structure (solid lines) and MoS$_{2}$ (dashed lines) absorption spectra with different structural parameters when $P = 480 nm$. (a) Using different radius $R$ of air holes, (b) using various PMMA layer and (c) SiO$_{2}$ layer thicknesses. The peak absorption (red lines) and peak wavelength (black lines) as functions of (d) $R$, (e) $d_p$  and (f) $d_s$. Other geometric parameters are consistent with Fig. 1.}
\end{figure}
As mentioned above, we have used the guided resonance to realize critical coupling and achieved the ultra-high-efficiency light absorption of monolayer MoS$_{2}$ in a MPA. We subsequently turn to investigate the dependence of the critical coupling on the structure parameters of the MPA. In the wavelengths of interest, the inherent loss of the structure mainly comes from MoS$_{2}$, whose absorption is largely independent of the wavelength as shown in Fig. 2(a). Consequently, for the design principle of critical coupling, we can concentrate on controlling the external leakage rate of the structure, mainly through the following three parameters. First of all, as shown in Fig. 4(a) and 4(d), the influence of the air holes radius $R$ on the absorption spectra of a MoS$_{2}$ monolayer (dashed lines) and the whole MPA (solid lines) is presented. From Fig. 4(a), we can see that even though the radius of the air hole is adjusted, the system is still in a critical coupling state, and the absorption peak of resonant wavelength exhibits a blue-shift with increasing the radius $R$ of air hole, which is due to the decrease of the effective refractive index of the structure with the expansion of the air hole. Moreover, as depicted in Fig. 4(d), we can also see that the influence on absorption peak (red line) and peak wavelength (black line) when the radius of air hole increases from $80$ nm to $280$ nm, where the peak wavelengths are linearly tuned by the radius of air hole, and the trend of absorption peak also shows that the system develops from undercoupling to critical coupling and then to over-coupling. Next, we consider the effect of the thickness of the PMMA layer on light absorption of monolayer MoS$_{2}$ (dashed lines) and the whole structure (solid lines). As shown in Fig. 4(b), compared with the previous parameter $R$, the thickness of PMMA layer is adjusted with the same spacing ($\Delta = 20$ nm), and the wavelength of the absorption peak is almost unchanged while the system is always in critical coupling. Fig. 4(e) shows the calculated peak absorption (red line) and peak wavelength (black line) as functions of the PMMA layer thickness. It has been found that the peak absorption of more than $99\%$ can be maintained when the PMMA layer thickness ($d_p$) ranges from $120$ nm to $280$ nm while other parameters are fixed as in Fig. 1. Meanwhile, the peak wavelength of the structure is almost invariable and maintain approximately linear relationship with the PMMA layer thickness. The main reason is that the external leakage rate ($\gamma_{e}$) of the structure is insensitive to the variation of the PMMA layer thickness in a certain range. We also consider the dependence of light absorption of monolayer MoS$_{2}$ and the entire structure on the thickness of SiO$_{2}$ layer in Fig. 4(c) and 4(f). We can find that when the thickness of SiO$_{2}$ layer is adjusted at the same distance, the absorption spectrum of the structure exhibits a significant red-shift with increasing the thickness of SiO$_{2}$ layer ($d_s$), and the absorption peak also changes slightly, as shown in Fig. 4(c). It can be seen in Fig. 4(f) that the absorption peak of the structure slowly rises up, and then drastically falls down when $d_s$ increases. The maximum appears in the critical coupling state of $d_s = 255$ nm, which indicates that the external leakage rate is sensitive to the change of the SiO$_{2}$ layer thickness. Meanwhile, the relationship between peak wavelength and thickness of SiO$_{2}$ layer is still nearly linear. To sum up, it can be seen that the designed structure has relatively large fabrication tolerances, and the tunable ultra-high-efficiency light absorption of monolayer MoS$_{2}$ in a MPA at the wavelengths of interest can be achieved by adjusting the above three parameters properly. 

\begin{figure}[htbp]
\centering
\includegraphics[scale=0.52]{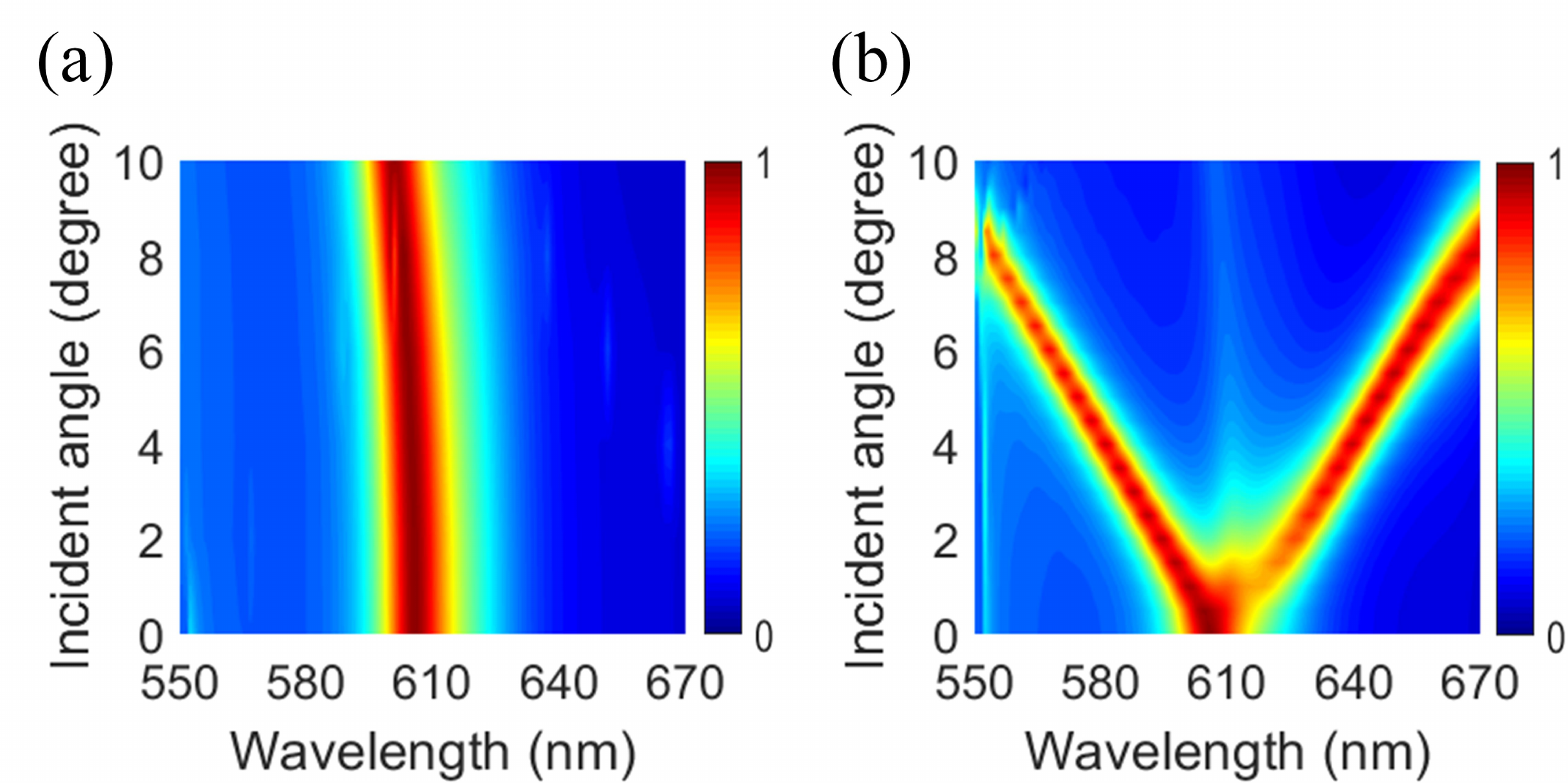}
\caption{Light absorption of monolayer MoS$_{2}$ as functions of incident angles $\theta$ for (a) TM-polarized and (b) TE-polarized light. The geometric parameters are the same as Fig. 1.}
\end{figure}
In addition, it is easy to find that the structure we designed has the nice property of being polarization-independent under normal incidence, due to its rotational symmetry. Meanwhile, we further investigate the light absorption of monolayer MoS$_{2}$ with different incident angle for TM-polarization and TE-polarization, respectively. From Fig. 5(a), we can find that the absorption of MoS$_{2}$ for TM-polarization remains almost invariant when the angle of incidence varies from $0^{\circ}$ to $10^{\circ}$, which is because the guided mode resonance is insensitive to the incident angle in this case\cite{fan2017monolayer}. In contrast, for TE-polarization, as shown in Fig. 5(b), the resonance exhibits wavelength splitting, and two major absorption peaks of the same degree appear in the absorption spectrum under oblique incidence, which forming multiple resonance. The two major absorption peaks originate from the excitation of other guided mode resonances, and they are sensitive to the incident angle, which is an important consideration in multispectral light detection\cite{piper2014total}. 

\begin{figure}[htbp]
\centering
\includegraphics[scale=0.32]{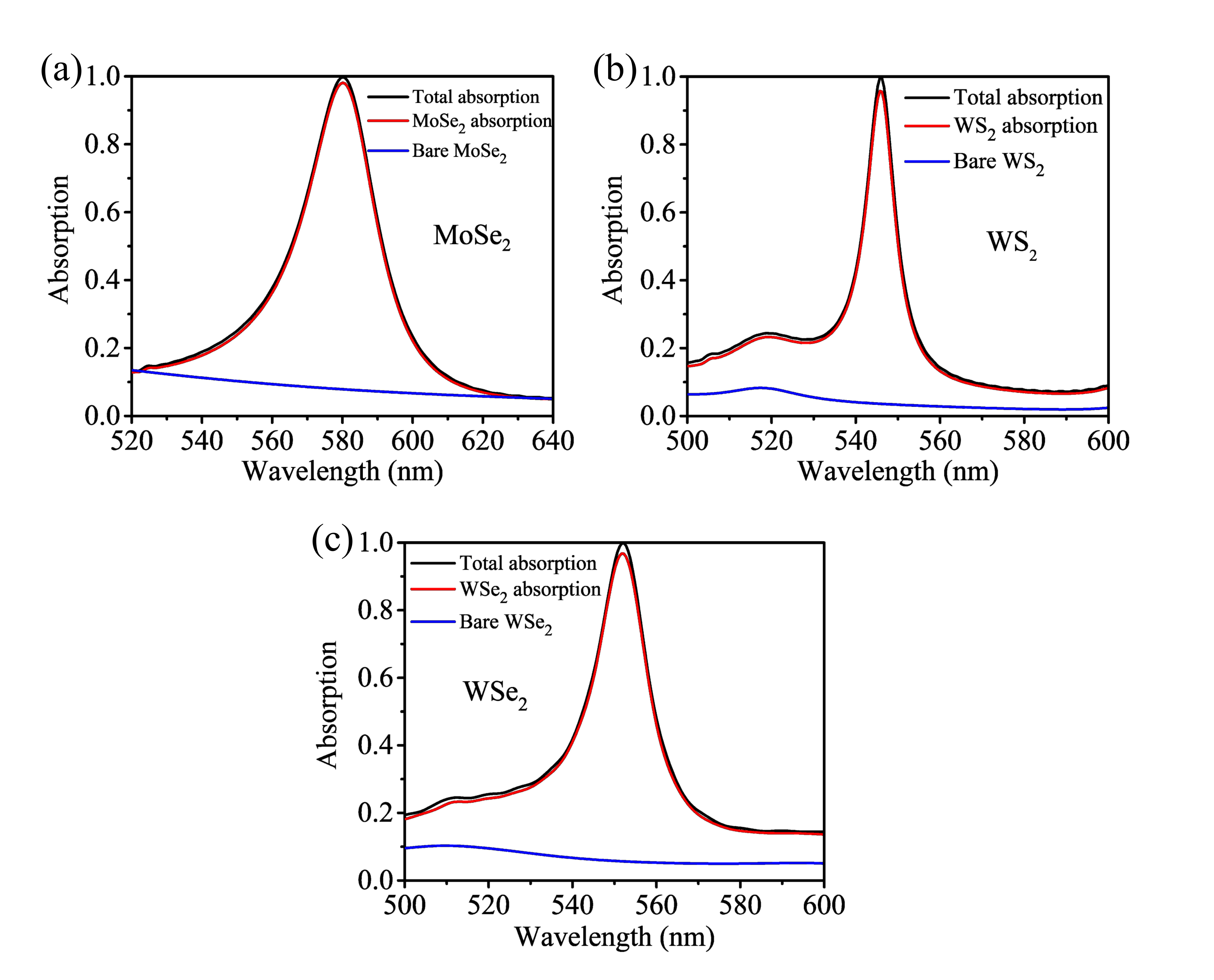}
\caption{The concept of critical coupling can be extended to other TMDCs, such as (a) MoSe$_{2}$, (b) WS$_{2}$ and (c) WSe$_{2}$. Absorption spectra of the whole MPA (black lines) and the monolayer TMDCs (red lines) with different thicknesses are compared with bare TMDCs monolayer suspended in air (blue lines) for $P = 480$ nm and $D = 200$ nm.}
\end{figure}
Finally, we also consider the general effects of the proposed critical coupling scheme on light absorption of other monolayer TMDCs materials, such as MoSe$_{2}$, WS$_{2}$, and WSe$_{2}$. FDTD simulations show that the absorption spectra of the designed TMDC-based perfect absorption system with different structural parameters under normal incidence, in which the monolayer MoS$_{2}$ is replaced by the monolayer MoSe$_{2}$, WS$_{2}$, and WSe$_{2}$, respectively. The total absorption of the structure (black lines) and the contribution from monolayer TMDCs (red lines), compared to the absorption of bare TMDCs stand in air (blue lines), as shown in Fig. 6. In addition, in the numerical simulations, the thickness of monolayer MoSe$_{2}$, WS$_{2}$, and WSe$_{2}$ are set to $0.646$ nm, $0.618$ nm, and $0.649$ nm, respectively. And the corresponding dielectric constant are derived from the experimental measurement of Li \textit{et al}\cite{li2014measurement}. For comparison, structure $1$, structure $2$, and structure $3$ represent a MPA with monolayer MoSe$_{2}$ (for $R = 200$ nm, $d_p = 205$ nm, and $d_s = 160$ nm) in Fig. 6(a), a MPA with monolayer WS$_{2}$ (for $R = 205$ nm, $d_p = 200$ nm, and $d_s = 100$ nm) in Fig. 6(b), and a MPA with monolayer WSe$_{2}$ (for $R = 210$ nm, $d_p = 210$ nm, and $d_s = 115$ nm) in Fig. 6(c), respectively. Absorption spectra corresponding to the above three structures show that the absorption of light in monolayer TMDCs can significantly reach $98.1\%$, $96\%$, and $96.8\%$, respectively. And the corresponding absorption peaks appear at $580.1$ nm, $545.7$ nm, and $551.7$ nm, respectively. Certainly, we can also change the operating wavelength by adjusting the structure parameters to achieve the wavelength selectivity of the system, which is the same as that of MoS$_{2}$. Therefore, the proposed TMDC-based perfect absorption system with its unique design principle can be applied to boost the absorption of other $2$D monolayer materials. 

\section{Conclusions}
A simple multilayer architecture with MoS$_{2}$ embedded in a MPA is proposed and its light absorption properties are numerically investigated. Absorption above $98.3\%$ in monolayer MoS$_{2}$ is obtained, which is over $12$ times more than that of bare monolayer MoS$_{2}$. The absorption enhancement of monolayer MoS$_{2}$ is mainly due to the strong confinement of electromagnetic field through the principle of critical coupling with guided resonance and the suppression of the loss caused by reflection and transmission in MPA. The operating wavelength can be modulated by adjusting the radius of air hole and the thickness of the PMMA and SiO$_{2}$ layers, which are valuable for the tunability and selectivity of the high efficiency absorption of monolayer MoS$_{2}$. Furthermore, the polarization independence of the structure at normal incidence and the angular characteristics of the system at oblique incidence are discussed by simulations. The geometry proposed here is general and can be extended with other monolayer TMDCs, such as MoSe$_{2}$, WS$_{2}$, and WSe$_{2}$. Our investigation provides guidance for the device design of using critical coupling theory to improve the optical response of monolayer TMDCs, and our results may be of great significance for TMDC-based applications, such as photodetection and photoluminescence.

\section*{Funding}
This work is supported by the National Natural Science Foundation of China (Grant No. 61376055 and 61775064), and the Fundamental Research Funds for the Central Universities (HUST: 2016YXMS024).
\end{document}